\documentclass[12pt,a4paper]{article}
\usepackage[dvips]{epsfig}

\input{tcilatex}
\input tcilatex

\begin{document}

\title{Bounding the tau neutrino magnetic moment from the process $e^{+}e^{-}$ $%
\rightarrow $ $\gamma \nu \overline{\nu }$ }
\author{Aytekin AYDEMIR\thanks{%
Mersin Universitesi, Icel- Turkey} \\
Ramazan SEVER\thanks{%
Orta Dogu Teknik Universitesi, Ankara- Turkey}}
\date{\today }

\thispagestyle{empty} 

\thispagestyle{empty}

\begin{titlepage}
\maketitle
\begin{abstract}
In a class of $E_6$ inspired models with a light additional neutral vector boson, we discuss the
effects of the magnetic moment of the tau neutrino by analyzing
$e^{+}e^{-}$ $\rightarrow $ $\gamma \nu \overline{\nu }$ at the $Z$-pole.
We take into account present scenarios for the extra Abelian group and
vary extra neutral gauge boson mass beyond the present experimental
exclusion limits. The present LEP experimental results prove us to set
an upper bound on the magnetic moment of the tau neutrino:
$\kappa _{\nu _{\tau }}$ $ \leq $ 1.83*10$^{-6}$ .
\end{abstract}
\end{titlepage}
\newpage

\section{INTRODUCTION}

Neutrinos are weakly interacting particles. They have very long mean free
path. So they immediately escape from the star and effect the evolution of
the star drastically. In the detection of neutrinos emerging from the sun,
the experimental results do not agree with the expected rates ($\sim $ 1/3
of the expected rates) [1] obtained from theoretical calculations using
standard model (SM). Hence this is considered as a solar neutrino problem.
On the other hand, due to the emission of neutrinos, some of the stars could
not be detected in their proper places [2]. Magnetic moment of the neutrino
is one of the properties to solve these problems. Magnetic moment can be the
reason of the deflection of the neutrinos by the magnetic field of the sun.

The possible electromagnetic properties of a massive Dirac neutrino are
summarized in the current [3,4,5]

\begin{equation}
J_{\mu }=\overline{u}_{v}(q_{2})[i\frac{e}{2m_{\nu }}%
F_{2}(q^{2})+eF_{3}(q^{2})\gamma _{5}]\sigma _{\mu \nu }q^{\nu }u_{v}(q_{1})
\end{equation}

where $F_{2}$ and $F_{3}$ are dimensionless structure functions,
corresponding to magnetic moment and electric dipole moment respectively at $%
q^{2}=0$.

On the other hand, some of the experimental results for the cross section of
some processes at high energies deflect from the values obtained
theoretically using standard model. One example is $e^{+}e^{-}\rightarrow
\mu ^{+}\mu ^{-}$ scattering cross section calculated at TRISTAN and LEP-I
energies and the result is 2$\sigma $ different from the theoretical
calculations [6,7]. This discrepancy could be removed if a new intermediate
vector boson (IVB) which comes out in the extension of the standard model or
in GUT's is considered [8,9,10]. In the successive breaking of $E_{6}$ as

$E_{6}->SO(10)XU(1)_{\psi }->SU(5)XU(1)_{\chi }XU(1)_{\psi }$

the standard elektroweak group $SU(3)_{c}XSU(2)_{L}XU(1)_{Y}$ is embedded in
$SU(5)$ subgroup of $SO(10).$ Therefore, we can write Q$^{^{\prime }}$
charges in terms of $Q_{\chi }$ and $Q_{\psi }$ which are already orthagonal
to symmetries of elektromagnetism and other known interactions which are
buried into $SU(5)$. Then we write [8,11]

\begin{eqnarray}
Q^{^{\prime }} &=&Q_{\chi }\cos \vartheta +Q_{\psi }\sin \vartheta .
\nonumber \\
Q^{^{\prime \prime }} &=&-Q_{\chi }\sin \vartheta +Q_{\psi }\cos \vartheta .
\end{eqnarray}

This new IVB Z$_{\vartheta }$ is considered in our calculations together
with the magnetic moment of the neutrino.

Thus, the relevant neutral current Lagrangian becomes [12,13]

\begin{equation}
-L_{NC}=g_1Z_0^\mu J_{Z_0\mu }+g_2Z_\vartheta ^\mu J_{Z_\vartheta \mu }
\end{equation}

with the currents

\begin{eqnarray}
J_{Z_0\mu } &=&\sum_f\overline{f}\gamma _\mu [a+b\gamma _5]f  \nonumber \\
J_{Z_\vartheta \mu } &=&\sum_f\overline{f}\gamma _\mu [a^{^{\prime
}}+b^{^{\prime }}\gamma _5]f
\end{eqnarray}

where $f$ representing fermions,

\begin{eqnarray}
g_1 &=&(g^2+g^{\prime 2})^{1/2}=\frac e{2\sin \theta _W\cos \theta _W}=(%
\sqrt{2}G_\mu M_{Z_0}^2)^{1/2}  \nonumber \\
g_2 &=&g_\vartheta
\end{eqnarray}

\begin{eqnarray}
a &=&-\frac{1}{2}+2\sin ^{2}\vartheta _{w}  \nonumber \\
b &=&\frac{1}{2}
\end{eqnarray}

\begin{eqnarray}
a^{^{\prime }} &=&X\cdot (\frac{\cos \vartheta }{\sqrt{6}}+\frac{\sin
\vartheta }{\sqrt{10}})\cdot (-\frac{\cos \vartheta }{\sqrt{6}}+\frac{3\sin
\vartheta }{\sqrt{10}})  \nonumber \\
b^{^{\prime }} &=&2\cdot X\cdot \frac{\sin \vartheta }{\sqrt{10}}(-\frac{%
\cos \vartheta }{\sqrt{6}}+\frac{3\sin \vartheta }{\sqrt{10}})
\end{eqnarray}

and

\begin{equation}
X=(\frac{g_{\vartheta }^{2}}{g^{2}+g^{^{\prime }2}})\cdot (\frac{M_{Z_{0}}}{%
M_{Z_{\vartheta }}})^{2}
\end{equation}

is a parameter depending on the coupling constant $g_\vartheta $ and the
mass of Z$_\vartheta ,$ where $\vartheta $ is the mixing angle in E$_6$

\begin{equation}
Z_{\vartheta }=Z_{\psi }\cos \vartheta +Z_{\chi }\sin \vartheta .
\end{equation}

For the following $\vartheta $ values the corresponding models emerge:

$\vartheta =0^{\circ },$ $Z_\vartheta \rightarrow Z_{\psi ,}$

$\vartheta =37.8^{\circ },$ $Z_\vartheta \rightarrow Z^{^{\prime }},$

$\vartheta =90^{\circ },$ $Z_\vartheta \rightarrow Z_{\chi ,}$

$\vartheta =127.8^{\circ },$ $Z_\vartheta \rightarrow Z_I$ [7].

\section{CALCULATION}

We calculated the differential cross section for the scattering process
\begin{equation}
e^{+}(p_1)+e^{-}(p_2)\rightarrow \gamma (k)+\nu (q_1)+\overline{\nu }(q_2)
\end{equation}

at LEP energies where only the relevant four Feynman diagrams are considered
and given in Fig. 1. Using the experimental result on the cross section of
abovementioned process we obtain an upper bound for the magnetic moment of
the tau neutrino ($\nu _{\tau }$). Charged currents can produce only $\nu
_{e}$ and $\nu _{\mu }$ due to lepton number conservation, and if considered
in the calculations, can only decrease the magnetic moment of the tau
neutrino. Therefore charged current contribution is not considered. The same
process is calculated at center of mass energies where $\sqrt{s}\leq
M_{Z_{0}}$ by D. Fargion et al. for six Feynman diagrams [14].

The four momenta of the electron, positron, photon, neutrino, and
antineutrino are $p_1$, $p_2$, $k$, $q_1$, and $q_2,$ respectively. The four
momenta of $Z_0$ and $Z_\vartheta $ are $s_1,$ and $s_2,$ respectively and
are taken equal to center of mass energy $\sqrt{s}$.

The total matrix element for the diagrams in Fig. 1, is

\begin{equation}
M_{Tot}=M_a+M_b+M_a^{^{\prime }}+M_b^{^{\prime }}
\end{equation}

where
\begin{eqnarray}
M_{a} &=&\overline{v}_{e^{+}}(p_{2})\frac{-ie}{\sin \vartheta _{w}\cos
\vartheta _{w}}\gamma _{\mu }(\frac{a+b\gamma _{5}}{2})u_{e^{-}}(p_{1})
\nonumber \\
&&\frac{i}{s-M_{Z_{0}}^{2}+i(s/M_{Z_{0}}^{2})M_{Z_{0}}\Gamma _{Z_{0}}}%
(-g_{\mu \nu }+\frac{s_{\mu }s_{\nu }}{M_{Z_{0}}^{2}})\cdot  \nonumber \\
&&\overline{u}_{v}(q_{1})\mu _{B}\sigma _{\alpha \beta }\epsilon ^{\alpha
}k^{\beta }(F_{2}+F_{3}\gamma _{5})\frac{i}{\not{q}_{1}+\not{k}\,}\left[
\frac{-ie}{\sin \vartheta _{w}\cos \vartheta _{w}}\gamma _{\nu }(\frac{%
1-\gamma _{5}}{4})\right] v_{\bar{\nu}}(q_{2})
\end{eqnarray}

\begin{eqnarray}
M_{b} &=&\overline{v}_{e^{+}}(p_{2})\frac{-ie}{\sin \vartheta _{w}\cos
\vartheta _{w}}\gamma _{\mu }(\frac{a+b\gamma _{5}}{2})u_{e^{-}}(p_{1})
\nonumber \\
&&\frac{i}{s-M_{Z_{0}}^{2}+i(s/M_{Z_{0}}^{2})M_{Z_{0}}\Gamma _{Z_{0}}}%
(-g_{\mu \nu }+\frac{s_{\mu }s_{\nu }}{M_{Z_{0}}^{2}})\cdot  \nonumber \\
&&\overline{u}_{v}(q_{1})\left[ \frac{-ie}{\sin \vartheta _{w}\cos \vartheta
_{w}}\gamma _{\nu }(\frac{1-\gamma _{5}}{4})\right] \frac{i}{\not{q}_{2}+%
\not{k}\,}\mu _{B}\sigma _{\alpha \beta }\epsilon ^{\alpha }k^{\beta
}(F_{2}+F_{3}\gamma _{5})v_{\overline{v}}(q_{2})
\end{eqnarray}

and for M$_a^{^{\prime }}$ and M$_b^{^{\prime }}$%
\begin{eqnarray}
M_a^{^{\prime }} &=&M_a(a\rightarrow a^{^{\prime }},b\rightarrow b^{^{\prime
}},M_{Z_0}\rightarrow M_{Z_\vartheta }) \\
M_b^{^{\prime }} &=&M_b(a\rightarrow a^{^{\prime }},b\rightarrow b^{^{\prime
}},M_{Z_0}\rightarrow M_{Z_\vartheta })
\end{eqnarray}

\begin{equation}
|M|^{2}=|M_{a}|^{2}+|M_{b}|^{2}+|M_{a}^{^{\prime }}|^{2}+|M_{b}^{^{\prime
}}|^{2}+(M_{a}M_{a}^{^{\prime }\dagger }+M_{b}M_{b}^{^{\prime }\dagger
}+M_{a}^{\dagger }M_{a}^{^{\prime }}+M_{b}^{\dagger }M_{b}^{^{\prime }}).
\end{equation}

Here only the terms different from zero are written. In Eq. (16), the first
two terms are due to the intermediate vector boson Z$_{0}$ and are
calculated by T. M. Gould and I. Z. Rothstein [4]. The second two terms are
due to Z$_{\vartheta }$ which arise in the extensions of the standard model.
The last four terms in parentheses are due to the mixing of Z$_{0}$ and Z$%
_{\vartheta }$. We calculate the last six terms. In the following equations $%
\Gamma _{Z_{i}}$ should be replaced by $\Gamma _{Z_{i}}\rightarrow \frac{s}{%
M_{Z_{i}}^{2}}\Gamma _{Z_{i}}$ where $i=0$ and $\vartheta .$

Taking the sum over final particle's polarizations and the average over the
initial particle's polarizations, the square of the matrix elements becomes

\begin{eqnarray}
\frac{1}{4}\sum\limits_{pol,spins}|M_{a}|^{2} &=&\frac{1}{4}\cdot \frac{%
e^{4}\mu _{B}^{2}\epsilon ^{\alpha }\epsilon ^{\xi }k^{\beta }k^{\eta }}{%
64\cdot \sin ^{4}\vartheta _{w}\cos ^{4}\vartheta
_{w}(q_{2}+k)^{4}((s-M_{Z_{0}}^{2})^{2}+M_{Z_{0}}^{2}\Gamma _{Z_{0}}^{2})}%
\cdot  \nonumber \\
&&[Tr[(\not{p}_{2}-m)\gamma _{\mu }(a+b\gamma _{5})(\not{p}_{1}+m)\gamma
_{\nu }(a+b\gamma _{5})]\cdot  \nonumber \\
&&Tr[\not{q}_{2}\gamma ^{\nu }(1-\gamma _{5})(\not{q}_{1}+\not{k}%
)(F_{2}-F_{3}\gamma _{5})\sigma _{\xi \eta }\not{q}_{1}\sigma _{\alpha \beta
}(F_{2}+F_{3}\gamma _{5})\cdot  \nonumber \\
&&(\not{q}_{1}+\not{k})\gamma ^{\mu }(1-\gamma _{5})]]
\end{eqnarray}

and

\begin{eqnarray}
\frac 14\sum\limits_{pol,spins}|M_b|^2 &=&\frac 14\cdot \frac{e^4\mu
_B^2\epsilon ^\alpha \epsilon ^\xi k^\beta k^\eta }{64\cdot \sin ^4\vartheta
_w\cos ^4\vartheta _w(q_2+k)^4[(s-M_{Z_0}^2)^2+M_{Z_0}^2\Gamma _{Z_0}^2]}%
\cdot  \nonumber \\
&&[Tr[(\not{p}_2-m)\gamma _\mu (a+b\gamma _5)(\not{p}_1+m)\gamma _\nu
(a+b\gamma _5)]\cdot  \nonumber \\
&&Tr[\not{q}_2(F_2-F_3\gamma _5)\sigma _{\xi \eta }(\not{q}_2+\not{k})\gamma
^\nu (1-\gamma _5)\not{q}_1\gamma ^\mu (1-\gamma _5)\cdot  \nonumber \\
&&(\not{q}_2+\not{k})\sigma _{\alpha \beta }(F_2+F_3\gamma _5)]].
\end{eqnarray}

The trace calculations yield

\begin{eqnarray}
MK1 &=&\frac 14\sum\limits_{pol,spins}(|M_a|^2+|M_b|^2)=\frac{\frac 14\cdot
16\cdot \kappa ^2\cdot e^4\cdot \mu _B^2}{8\sin ^4\vartheta _w\cos
^4\vartheta _w[(s-M_{Z_0}^2)^2+M_{Z_0}^2\Gamma _{Z_0}^2]}\cdot  \nonumber \\
&&[4(p_1.q_1)(p_2.q_2)\sin ^4\vartheta _w+4(p_1.q_2)(p_2.q_1)\sin
^4\vartheta _w-  \nonumber \\
&&4(p_1.q_2)(p_2.q_1)\sin ^2\vartheta _w+(p_1.q_2)(p_2.q_1)],
\end{eqnarray}

\begin{eqnarray}
MK2 &=&\frac 14\sum\limits_{pol,spins}(|M_a^{^{\prime }}|^2+|M_b^{^{\prime
}}|^2)=\frac{\frac 14\cdot 16\cdot \kappa ^2\cdot e^4\cdot \mu _B^2\cdot X^2%
}{7200\cdot \sin ^4\vartheta _w\cos ^4\vartheta _w\cdot [(s-M_{Z_\vartheta
}^2)^2+M_{Z_\vartheta }^2\Gamma _{Z_\vartheta }^2]}\cdot  \nonumber \\
&&[729(p_1.q_1)(p_2.q_2)\sin ^4\vartheta -270\cdot (p_1.q_1)(p_2.q_2)\sin
^2\vartheta \cos ^2\vartheta _{}+  \nonumber \\
&&25(p_1.q_1)(p_2.q_2)\cos ^4\vartheta -72\cdot \sqrt{15}(p_1.q_2)(p_2.q_1)%
\sin ^3\vartheta \cos \vartheta -  \nonumber \\
&&40\cdot \sqrt{15}(p_1.q_2)(p_2.q_1)\sin \vartheta \cos ^3\vartheta +
\nonumber \\
&&81(p_1.q_2)(p_2.q_1)\sin ^4\vartheta +330(p_1.q_2)(p_2.q_1)\sin
^2\vartheta \cos ^2\vartheta +  \nonumber \\
&&25(p_1.q_2)(p_2.q_1)\cos ^4\vartheta ],
\end{eqnarray}

\begin{eqnarray}
MK3 &=&\frac 14\sum\limits_{pol,spins}(M_aM_a^{^{\prime }\dagger
}+M_bM_b^{^{\prime }\dagger }+M_a^{\dagger }M_a^{^{\prime }}+M_b^{\dagger
}M_b^{^{\prime }})=  \nonumber \\
&&\frac{\frac 14\cdot \frac{16}{240}\cdot \kappa ^2\cdot e^4\cdot \mu
_B^2\cdot X}{\sin ^4\vartheta _w\cos ^4\vartheta _w}\cdot f(s,M_{Z_0},\Gamma
_{Z_0},M_{Z_{_\vartheta }},\Gamma _{Z_{_\vartheta }})\cdot  \nonumber \\
&&[54\cdot (p_1.q_1)(p_2.q_2)\sin ^2\vartheta _w\sin ^2\vartheta -10\cdot
(p_1.q_1)(p_2.q_2)\sin ^2\vartheta _w\cos ^2\vartheta +  \nonumber \\
&&8\cdot \sqrt{15}\cdot (p_1.q_2)(p_2.q_1)\sin ^2\vartheta _w\sin \vartheta
\cos \vartheta -4\cdot \sqrt{15}\cdot (p_1.q_2)(p_2.q_1)\sin \vartheta \cos
\vartheta -  \nonumber \\
&&18\cdot (p_1.q_2)(p_2.q_1)\sin ^2\vartheta _w\sin ^2\vartheta -10\cdot
(p_1.q_2)(p_2.q_1)\sin ^2\vartheta _w\cos ^2\vartheta +  \nonumber \\
&&9\cdot (p_1.q_2)(p_2.q_1)\sin ^2\vartheta +5\cdot (p_1.q_2)(p_2.q_1)\cos
^2\vartheta ].
\end{eqnarray}

where

$f(s,M_{Z_{0}},\Gamma _{Z_{0}},M_{Z_{_{\vartheta }}},\Gamma _{Z_{_{\vartheta
}}})=\frac{2([s-M_{Z_{0}}^{2}]\cdot \lbrack s-M_{Z_{\vartheta
}}^{2}]+M_{Z_{0}}\Gamma _{Z_{0}}\cdot M_{Z_{_{\vartheta }}}\Gamma
_{Z_{_{\vartheta }}})}{\{[s-M_{Z_{0}}^{2}]\cdot \lbrack s-M_{Z_{\vartheta
}}^{2}]+M_{Z_{0}}\Gamma _{Z_{0}}\cdot M_{Z_{_{\vartheta }}}\Gamma
_{Z_{_{\vartheta }}})\}^{2}+[(s-M_{Z_{\vartheta }}^{2})\cdot M_{Z_{0}}\Gamma
_{Z_{0}}-(s-M_{Z_{0}}^{2})\cdot M_{Z_{_{\vartheta }}}\Gamma _{Z_{_{\vartheta
}}}]^{2}}.$

\bigskip

Since we are interested with the magnetic moment in all of the above
calculations we considered F$_{3}=0$ and F$_{2}=\kappa $ and $m_{e}^{2}\ll
s. $

Now, using the differential cross section

\begin{equation}
d\sigma =\frac{|M|^2}{4[(p_1.p_2)^2-m_em_e]^{1/2}}\frac{d^3q_1}{2q_{10}(2\pi
)^3}\frac{d^3q_2}{2q_{20}(2\pi )^3}\frac{d^3k}{2k_{0}(2\pi )^3}(2\pi
)^4\delta ^4(p_1+p_2-q_1-q_2-k)
\end{equation}

with Lenard's formula

\begin{equation}
\int \frac{d^3q_1}{2q_{10}}\frac{d^3q_2}{2q_{20}}q_{1\mu }q_{2\nu }\delta
^4(P-q_1-q_2)=\frac \pi {24}[2P_\mu P_\nu +g_{\mu \nu }P^2],
\end{equation}

we obtain

\begin{eqnarray}
d\sigma &=&d(\sigma _{1}+\sigma _{2}+\sigma _{3})=E_{\gamma }dE_{\gamma
}d(\cos \vartheta _{\gamma })\cdot \kappa ^{2}\alpha ^{2}\mu _{B}^{2}\cdot
\nonumber \\
&&\{-\frac{1}{96\pi \sin ^{4}\vartheta _{w}\cos ^{4}\vartheta _{w}\cdot
\lbrack (s-M_{Z_{0}}^{2})^{2}+M_{Z_{0}}^{2}\Gamma _{Z_{0}}^{2}]}  \nonumber
\\
&&[4\sqrt{s}(8\sin ^{4}\vartheta _{w}-4\sin ^{2}\vartheta _{w}+1)E_{\gamma }+
\nonumber \\
&&4(2\sin ^{2}\vartheta _{w}-1)(\cos ^{2}\vartheta _{\gamma }E_{\gamma
}^{2}-2s-E_{\gamma }^{2})\sin ^{2}\vartheta _{w}+\cos ^{2}\vartheta _{\gamma
}E_{\gamma }^{2}-2s-E_{\gamma }^{2}]+  \nonumber \\
&&\frac{X^{2}}{43200\cdot \pi \sin ^{4}\vartheta _{w}\cos ^{4}\vartheta
_{w}\cdot \lbrack (s-M_{Z_{\vartheta }}^{2})^{2}+M_{Z_{\vartheta
}}^{2}\Gamma _{Z_{\vartheta }}^{2}]}\cdot  \nonumber \\
&&[4\sqrt{s}(4\sqrt{15}(9\sin ^{2}\vartheta +5\cos ^{2}\vartheta )\sin
\vartheta \cos \vartheta -405\sin ^{4}\vartheta -  \nonumber \\
&&30\sin ^{2}\vartheta \cos ^{2}\vartheta -25\cos ^{4}\vartheta )E_{\gamma
}-2((2s+E_{\gamma }^{2})-\cos ^{2}\vartheta _{\gamma }E_{\gamma }^{2})\cdot
(18\sqrt{15}\sin ^{2}\vartheta +  \nonumber \\
&&10\sqrt{15}\cos ^{2}\vartheta -15\sin \vartheta \cos \vartheta )\sin
\vartheta \cos \vartheta -  \nonumber \\
&&405(\cos ^{2}\vartheta _{\gamma }E_{\gamma }^{2}-2s-E_{\gamma }^{2})\sin
^{4}\vartheta +25(2s+E_{\gamma }^{2})\cos ^{4}\vartheta -25\cos
^{2}\vartheta _{\gamma }\cos ^{4}\vartheta E_{\gamma }^{2}]-  \nonumber \\
&&\frac{X}{2880\pi \sin ^{4}\vartheta _{w}\cos ^{4}\vartheta _{w}}\cdot
f(s,M_{Z_{0}},\Gamma _{Z_{0}},M_{Z_{_{\vartheta }}},\Gamma _{Z_{_{\vartheta
}}}).  \nonumber \\
&&[4\sqrt{s}(4\sqrt{15}(2\sin ^{2}\vartheta _{w}-1)\sin \vartheta \cos
\vartheta +4(9\sin ^{2}\vartheta -5\cos ^{2}\vartheta )  \nonumber \\
&&\sin ^{2}\vartheta _{w}+9\sin ^{2}\vartheta +5\cos ^{2}\vartheta
)E_{\gamma }+4[9(\cos ^{2}\vartheta _{\gamma }E_{\gamma }^{2}-2s-E_{\gamma
}^{2})\sin ^{2}\vartheta +  \nonumber \\
&&5(2s+E_{\gamma }^{2})\cos ^{2}\vartheta -5\cos ^{2}\vartheta _{\gamma
}\cos ^{2}\vartheta E_{\gamma }^{2}]\sin ^{2}\vartheta _{w}+  \nonumber \\
&&4\sqrt{15}(2\sin ^{2}\vartheta _{w}-1)(\cos ^{2}\vartheta _{\gamma
}E_{\gamma }^{2}-2s-E_{\gamma }^{2})\sin \vartheta \cos \vartheta +
\nonumber \\
&&9(\cos ^{2}\vartheta _{\gamma }E_{\gamma }^{2}-2s-E_{\gamma }^{2})\sin
^{2}\vartheta -5(2s+E_{\gamma }^{2})\cos ^{2}\vartheta +5\cos ^{2}\vartheta
_{\gamma }\cos ^{2}\vartheta E_{\gamma }^{2}]\}.
\end{eqnarray}

Integrating over $\vartheta _\gamma $ from -$\pi $ to 0 and $E_\gamma $ from
15 GeV to 100 GeV and using the following numerical values, sin$^2$ $%
\vartheta _w=0.2314,$ $\mu _B=5.85*10^{-18}$ $GeV/Gauss$, $\alpha =\frac
1{137}$ , $M_{Z_0}=91.187$ $GeV,$ $s=M_{Z_0}^2,$ $X=\frac{1.15}3\cdot \left(
\frac{M_{Z_0}}{M_{Z_\vartheta }}\right) ^2,\Gamma _{Z_\vartheta }=\Gamma
_{Z_0}=2.49$ $GeV$,

we obtain the cross section

\begin{equation}
\sigma =\sigma (\kappa ,M_{Z_\vartheta },\vartheta ).
\end{equation}

Since we have calculated the cross section at the Z- pole, that is at $%
s=M_{Z_{0}}^{2},$ the value of sin$^{2}\vartheta _{w}$ is not effected by Z$%
_{\theta }$ physics [15,16], and we take sin$^{2}\vartheta _{w}=0.2314.$ And
also since $\Gamma _{Z_{_{\vartheta }}}$ contribution is small at $%
s=M_{Z_{0}}^{2},$ taking $\Gamma _{Z_{_{\vartheta }}}=\Gamma _{Z_{_{0}}}$ or
$\Gamma _{Z_{_{\vartheta }}}=0$ does not effect our results [7]. We have
calculated the cross section due to their magnetic moments for various
values of $\vartheta $ and $M_{Z_{\vartheta }}$ corresponding to different
models of new physics due to superstring-inspired E$_{6}$ theories [8] and
the numerical results for the cross sections for all neutrino interactions
are given in Table 1.

The cross sections $\sigma _{1},$ $\sigma _{2},$ $\sigma _{3},$ correspond
to $Z_{0}$ (SM), new intermediate vector boson $Z_{\vartheta },$ mixing of $%
Z_{0}\ $and $Z_{\vartheta }$ respectiveley$.$ The total cross section is $%
\sigma =\sigma _{1}+$ $\sigma _{2}+$ $\sigma _{3}.$

Using $\sigma _{\exp }L=N$ and taking the experimental results [17], L=48.0
pb$^{-1},$ N=14, we obtain a limit for the neutrino magnetic moment $\kappa $
in terms of $\mu _{B}$ as

$\kappa \leq 1.83*10^{-6}$

for $\sigma $ for $M_{Z_\vartheta }=7\cdot M_{Z_0},$ and $\vartheta =37.8.$

\newpage

\section{DISCUSSION AND CONCLUSION}

We have seen that, there is no much difference for various $M_{Z_{\vartheta
}}$ and $\vartheta $ values in the calculation of the cross section in the
content of this study. The only important contribution from $Z_{\vartheta }$
particle is at $\vartheta =37.8^{\circ }$ which corresponds to $Z_{\vartheta
}\rightarrow Z^{^{\prime }}.$ That is, it gives negative contribution to the
cross section.

We conclude that there is no further constraint on the magnetic moment of
the neutrino due to Z$_{\vartheta }$ since its mass is higher than Z$_{0}$
at $\sqrt{s}=M_{Z_{0}}.$ But at higher center-of-mass energies $\sqrt{s}\sim
M_{Z_{\vartheta }},$ the Z$_{\vartheta }$ contribution to the cross section
becomes comparable with Z$_{0}.$

ACKNOWLEDGMENT

We thank Professor T. Aliev for the suggestion of the problem, and also
thank Dr. D. A. Demir for helpful discussions.

\newpage

FIGURE CAPTIONS

FIGURE 1. The lowest order Feynman diagrams for the pair annihilation
process.

\newpage

TABLE CAPTIONS

TABLE 1. Scattering cross section for various $\vartheta $ and $%
M_{Z_\vartheta }$ values at $\sqrt{s}=$M$_{Z_0}.$ The unit of $\sigma $ is cm%
$^2.$

\begin{figure}[tbp]
\unitlength1mm
\begin{picture}(161,135)
\put(5,-3){\psfig{file=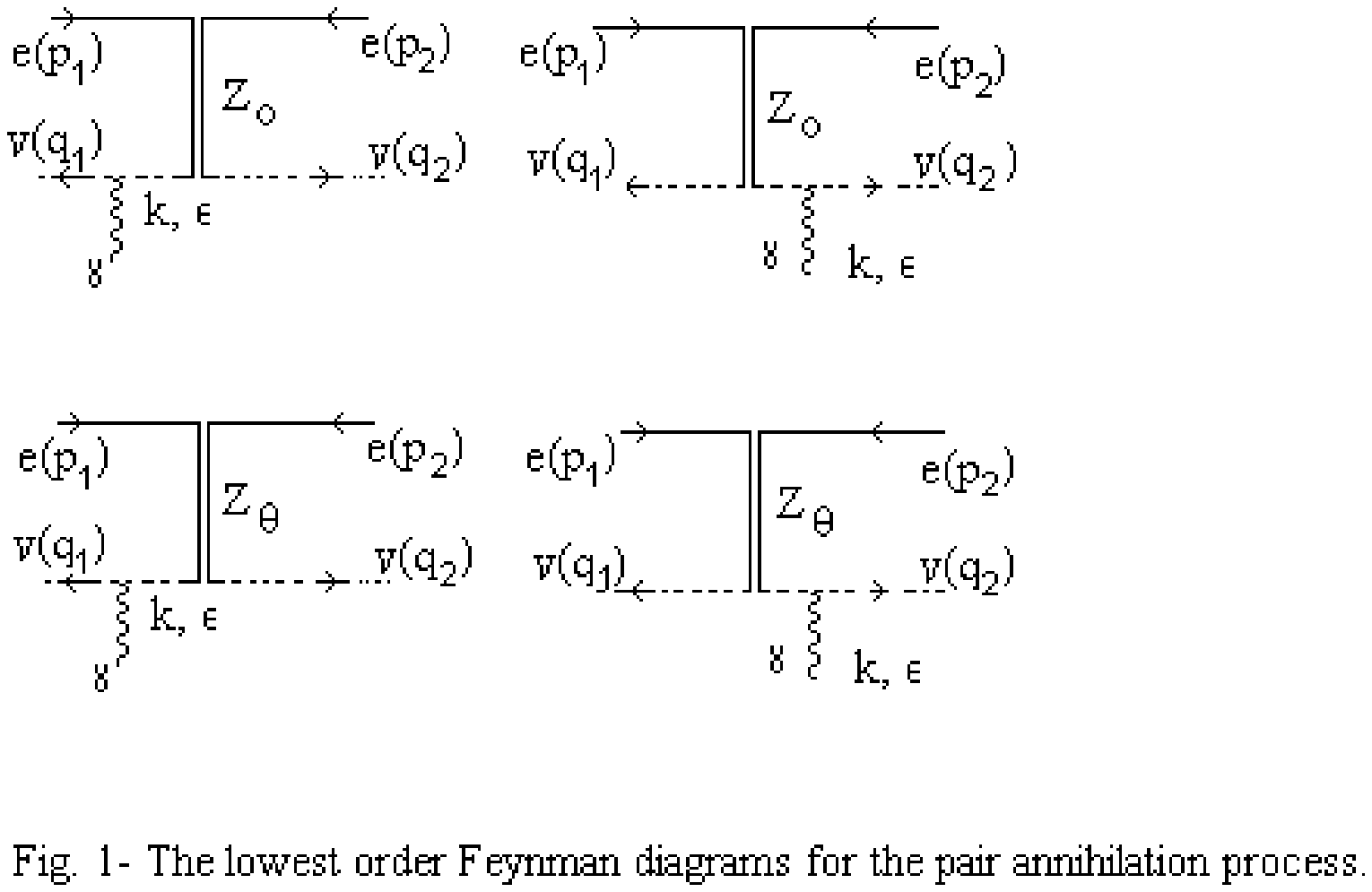,width=22cm} }
\end{picture}
\end{figure}

\newpage \ TABLE I. Scattering cross sections for various $\vartheta $ and $%
M_{Z_\vartheta }$ values at $\sqrt{s}=$M$_{Z_0}.$ The unit of $\sigma $ is cm%
$^2.$

$
\begin{array}[b]{ccccc}
&  & \vartheta =0 &  &  \\
\frac{M_{Z_\vartheta }}{M_{Z_0}} & \frac{\sigma _1}{\kappa ^2} & \frac{%
\sigma _2}{\kappa ^2} & \frac{\sigma _3}{\kappa ^2} & \frac \sigma {\kappa
^2} \\
5 & 8.682*10^{-26} & 2.913*10^{-36} & 1.696*10^{-35} & 8.682*10^{-26} \\
6 & 8.682*10^{-26} & 6.605*10^{-37} & 4.615*10^{-36} & 8.682*10^{-26} \\
7 & 8.682*10^{-26} & 1.896*10^{-37} & 1.545*10^{-36} & 8.682*10^{-26} \\
8 & 8.682*10^{-26} & 6.450*10^{-38} & 6.009*10^{-37} & 8.682*10^{-26} \\
9 & 8.682*10^{-26} & 2.497*10^{-38} & 2.617*10^{-37} & 8.682*10^{-26} \\
10 & 8.682*10^{-26} & 1.070*10^{-38} & 1.246*10^{-37} & 8.682*10^{-26} \\
11 & 8.682*10^{-26} & 4.974*10^{-39} & 6.372*10^{-38} & 8.682*10^{-26} \\
&  & \vartheta =37.8 &  &  \\
& \frac{\sigma _1}{\kappa ^2} & \frac{\sigma _2}{\kappa ^2} & \frac{\sigma _3%
}{\kappa ^2} & \frac{\sigma _{}}{\kappa ^2} \\
5 & 8.682*10^{-26} & 1.982*10^{-36} & -1.322*10^{-35} & 8.682*10^{-26} \\
6 & 8.682*10^{-26} & 4.495*10^{-37} & -3.597*10^{-36} & 8.682*10^{-26} \\
7 & 8.682*10^{-26} & 1.290*10^{-37} & -1.204*10^{-36} & 8.682*10^{-26} \\
8 & 8.682*10^{-26} & 4.390*10^{-38} & -4.683*10^{-37} & 8.682*10^{-26} \\
9 & 8.682*10^{-26} & 1.700*10^{-38} & -2.040*10^{-37} & 8.682*10^{-26} \\
10 & 8.682*10^{-26} & 7.281*10^{-39} & -9.710*10^{-38} & 8.682*10^{-26} \\
11 & 8.682*10^{-26} & 3.385*10^{-39} & -4.965*10^{-38} & 8.682*10^{-26}
\end{array}
$

$
\begin{array}[b]{ccccc}
&  & \vartheta =90 &  &  \\
\frac{M_{Z_\vartheta }}{M_{Z_0}} & \frac{\sigma _1}{\kappa ^2} & \frac{%
\sigma _2}{\kappa ^2} & \frac{\sigma _3}{\kappa ^2} & \frac \sigma {\kappa
^2} \\
5 & 8.682*10^{-26} & 3.675*10^{-35} & 7.869*10^{-34} & 8.682*10^{-26} \\
6 & 8.682*10^{-26} & 8.333*10^{-36} & 2.141*10^{-34} & 8.682*10^{-26} \\
7 & 8.682*10^{-26} & 2.391*10^{-36} & 7.170*10^{-35} & 8.682*10^{-26} \\
8 & 8.682*10^{-26} & 8.138*10^{-37} & 2.788*10^{-35} & 8.682*10^{-26} \\
9 & 8.682*10^{-26} & 3.151*10^{-37} & 1.214*10^{-35} & 8.682*10^{-26} \\
10 & 8.682*10^{-26} & 1.350*10^{-37} & 5.781*10^{-36} & 8.682*10^{-26} \\
11 & 8.682*10^{-26} & 6.275*10^{-38} & 2.956*10^{-36} & 8.682*10^{-26} \\
&  & \vartheta =127.8 &  &  \\
& \frac{\sigma _1}{\kappa ^2} & \frac{\sigma _2}{\kappa ^2} & \frac{\sigma _3%
}{\kappa ^2} & \frac{\sigma _{}}{\kappa ^2} \\
5 & 8.682*10^{-26} & 3.135*10^{-35} & 7.398*10^{-34} & 8.682*10^{-26} \\
6 & 8.682*10^{-26} & 7.108*10^{-36} & 2.013*10^{-34} & 8.682*10^{-26} \\
7 & 8.682*10^{-26} & 2.040*10^{-36} & 6.740*10^{-35} & 8.682*10^{-26} \\
8 & 8.682*10^{-26} & 6.942*10^{-37} & 2.621*10^{-35} & 8.682*10^{-26} \\
9 & 8.682*10^{-26} & 2.688*10^{-37} & 1.142*10^{-35} & 8.682*10^{-26} \\
10 & 8.682*10^{-26} & 1.151*10^{-37} & 5.435*10^{-36} & 8.682*10^{-26} \\
11 & 8.682*10^{-26} & 5.353*10^{-38} & 2.779*10^{-36} & 8.682*10^{-26}
\end{array}
$

\end{document}